# The common origin of family and non-family asteroids

*Stanley F. Dermott[1], Apostolos A. Christou[2], Dan Li,[3] Thomas. J. J. Kehoe[4] & J. Malcolm Robinson[1]

[1]University of Florida, Department of Astronomy, Gainesville, Florida 32611, USA.

[2]Armagh Observatory and Planetarium, College Hill, Armagh BT61 9DG, UK.

[3]University of Pennsylvania, Department of Physics and Astronomy, Philadelphia, PA 19104, USA.

[4]Florida Space Institute, 12354 Research Parkway, Orlando, FL 32826, USA

**\*Corresponding author**

Correspondence to Stanley F. Dermott

All asteroids are currently classified as either family, originating from the disruption of known bodies[1], or non-family. An outstanding question is the origin of these non-family asteroids. Were they formed individually, or as members of known families but with chaotically evolving orbits, or are they members of old, ghost families, that is, asteroids with a common parent body but with orbits that no longer cluster in orbital element space? Here we show that the sizes of the non-family asteroids in the inner belt are correlated with their orbital eccentricities and anticorrelated with their inclinations, suggesting that both non-family and family asteroids originate from a small number of large primordial planetesimals. We estimate that ∼ 85% of the asteroids in the inner main belt originate from the Flora, Vesta, Nysa, Polana and Eulalia families with the remaining ∼15% originating from either the same families or, more likely, a few ghost families. These new results imply that we must seek explanations for the differing characteristics of the various meteorite groups in the evolutionary histories of a few, large, precursor bodies[2]. Our findings also support the model that asteroids formed big through the gravitational collapse of material in a protoplanetary disk[3].

There are now over 400,000 asteroids in the main belt with numbered orbits[4] with about one third of these asteroids located in the inner main belt (IMB) defined as $2.1 < a < 2.5\ au$, where a is the proper (or long-term average) semimajor axis. The families in the IMB[5,6] defined by the Hierarchical Clustering Method[7] (HCM), are listed in

Supplementary Table 1. HCM is a clustering algorithm that defines the separation distance, $d_{cut}$, of orbits in $a - e - I$ space, where $e$ and $I$ are the proper eccentricity and the proper inclination. The choice of $d_{cut}$ is arbitrary and the confidence level in family membership decreases as $d_{cut}$ increases[6]. Nesvorný[5] lists five major families in the IMB (Vesta, Flora and the Nysa-Polana-Eulalia complex which we count as three families) and there are a further ~2,000 asteroids with $H < 16.5$ in the small Hungaria family interior to the IMB (Supplementary Table 1). Family members with $H < 16.5$ constitute 44% of the asteroids in the IMB and are one likely source of meteorites and near-Earth asteroids (NEAs)[2]. The only other likely sources in the IMB are the remaining 56% of asteroids now classified by Nesvorný[5,6] as non-family.

The distributions of the proper $e$ and $I$ of the asteroids in the IMB, as calculated by Kneževic and Milani[4] (listed in URL:http://hamilton.dm.unipi.it/~astdys2/propsynth/all.syn) are shown in Figure 1. While the discussion in this paper has application to all of the asteroids in the main belt, because the IMB is more observationally complete than the rest of the main belt, we confine our analysis here to the IMB. By comparing the latest, 2017 listing of asteroid orbital elements with the earlier, 2014 listing (Supplementary Figure 1), we estimate that the completeness limit of the IMB varies from absolute magnitude $H = 17$ close to the inner edge at 2.1 $au$, to $H = 16.5$ close to the outer edge at 2.5 $au$. Here, we use the conservative limit of $H = 16.5$ for the whole IMB, consistent with an earlier, 2008 estimate of the total number of multi- and single-opposition objects[8]. From Figure 1 and Supplementary Table 1, we see that the major families originate from a few large asteroids with quite different $e$ and $I$. Figure 1d shows the distribution of $e$ and $I$ of the non-family asteroids alone. The holes in this distribution are caused by the absence of the family asteroids[5]. The halos surrounding these holes appear to be associated with the major (most numerous) families, suggesting that the two populations are related.[6,9]

Plots of semimajor axis against $H$ or $1/D$, where $D$ is the asteroid diameter, are crucial diagnostics of the dynamical evolution of asteroid families. These plots reveal characteristic V-shaped distributions that can be accounted for by the action of Yarkovsky radiation forces[10]. In some cases, these plots have yielded estimates of the family ages[11]. Here we perform analogous tests for the non-family asteroids by analyzing the dependence of their mean proper $e$ and mean proper $I$ on $H$. Figure 2 shows that the separate families have negligible changes of mean $e$ and mean $I$ with $H$ even for $H > 16.5$, the completeness limit. This is because the $e$ and $I$ of the family members are, by definition, confined to the narrow ranges shown in Supplementary Table 1. However, the plots for the family asteroids as a whole show strong changes with $H$. For these asteroids we observe a decrease in mean $e$ and an increase in mean $I$ with decreasing $H$. For the non-family asteroids, we also observe strong changes of both mean $e$ and mean I with $H$, similar to those changes shown by the families. The statistical significances of the observed correlations are given in Supplementary Table 2.

We account for these observations through an analysis of the asteroid size-frequency distributions (SFDs) shown in Figure 3. While the SFDs of the non-family asteroids and the family asteroids as a whole are closely similar, apart from a small scaling factor, Figures 3b and 3d show that the SFDs of the separate families vary markedly. In Figure 3b, we fit cubic polynomials to the SFDs in the range $16.5 > H > 14$. In Figure 3c, we show how the slopes of the SFDs vary with $H$. In Figure 3d, we show ratios of the number of asteroids $dN$ in a box of width $dH = 0.1$ for the major families and for the separate family and non-family asteroids. It is striking that the best-fit curves for the change with $H$ of the ratio of all families to non-family is close to flat, in contrast to the large changes with $H$ observed for the ratio of Nysa-Polana-Eulalia to Vesta, the ratio of Nysa-Polana-Eulalia to Flora and the ratio of Vesta to Flora.

Two processes determine the SFDs of asteroids. Firstly, the large asteroids that suffered catastrophic collisions or major cratering events produced family members with an SFD determined by the size, strength and impact history of the precursor asteroid. Secondly, Dohnanyi[12] showed that if the impact strength of an asteroid is independent of its size, then, regardless of the SFD of the individual disruptions, after some considerable time the asteroid belt forms an equilibrium collisional cascade that can be described by a single SFD of the form

$$\log dN = bH + c \tag{1}$$

where $dN$ is the number of asteroids in a box of width $dH$ and $c$ is a constant for asteroids with the same albedo. Once an equilibrium cascade has been established, Dohnanyi[12] showed that $b = 0.5$. Numerical experiments[13] have shown that the relaxation timescale for the main belt exceeds $100 \ My$. Figure 3c shows that while the observed values of $b$ are close to 0.5 for $H \approx 14$, as $H$ tends to $\sim16.5, b$ tends to $\sim0.1$. This decrease in slope with increasing $H$ is common to all of the major families and because the asteroids in our set have $H < 16.5$, it cannot be due to observational selection. Any analysis of the evolution of the SFD needs to include the loss of asteroids due to Yarkovsky radiation forces[14,15], the disruption of small asteroids through YORP-induced rotational fission[16] and losses due to chaotic orbital evolution driven by point-mass gravitational forces[17,18]. These processes were not included in Dohnanyi's analysis[12].

The five largest families (Flora, Vesta and Nysa-Polana-Eulalia) dominate the SFD of the families as a whole and these families have quite different SFDs (Figure 3b). At high $H$ the SFD of the families as a whole is dominated by asteroids in the Nysa-Polana-Eulalia complex that have low $I$ and high $e$, whereas at low $H$ the SFD is dominated by asteroids in the Vesta family that have comparatively high $I$ and low $e$. The $e$ and $I$ of the Flora family are intermediate to those of the Vesta and Nysa-Polana-Eulalia families. We deduce from this that the changes of mean $e$ and mean $I$ with $H$ for the family asteroids as a whole are caused by the dominant family asteroids having both different SFDs and parent asteroids with distinctly different $e$ and $I$. This, in addition to the existence of the family halos (Figure 1d), plus the fact that the SFDs of the non-family asteroids and the family as a whole asteroids are closely similar (Figures 3b and 3d), suggests that the changes of mean $e$ and mean $I$ with $H$ for the non-family asteroids, which are also similar in shape to that of the family asteroids as a whole, arise because these non-family asteroids originate from the dominant major families in the IMB: Flora, Vesta and Nysa-Polana-Eulalia. This argument requires that the orbital elements of some family asteroids have dispersed, while the differences in their SFDs have survived.

If the non-family asteroids have evolved from the major families, then orbital evolution must have resulted in changes in both proper $e$ and proper $I$. Our numerical investigations have confirmed that chaotic orbital evolution in the IMB can result in significant dispersions on timescales less than the age of the solar system. We have also shown that the rate of orbital evolution of asteroids in mean motion resonances is greater than that of non-resonant asteroids, allowing the orbits of some family asteroids to disperse while the orbits of others remain more compact in $a - e - I$ space (Figure 4). Recently Delbo' et al.[19] detected a small, old family of large dark asteroids in the IMB. Because this family is small it does not influence our arguments here, but we do not rule out the presence of large, ghost families in the IMB.

Further progress in linking the non-family asteroids to the major families and eliminating or confirming the presence of old, ghost families may be aided in the future by an analysis of all the colour, albedo and spectral data. Here we strengthen our arguments by using the semimajor axes to separate the asteroids in the IMB into three groups having different dominant family members (Supplementary Figure 2). We also separate the non-family asteroids in the IMB into halo and non-halo asteroids (classified according to the criterion described in Supplementary Figure 3). Because of the arbitrary choice of $d_{cut}$, the division between family and non-family asteroids is already uncertain and dividing the non-family asteroids into halo and non-halo asteroids adds another layer of uncertainty. Here we justify this separation by showing that the halo and non-halo asteroids have quite different properties. While the mean inclinations and mean eccentricities of the family and halo asteroids are closely similar, the mean inclination of the non-halo asteroids is much greater than those of the family and the halo asteroids (Supplementary Figure 4). We also observe that the mean inclination of the non-halo asteroids increases with decreasing $H$, whereas the corresponding curves for the family and halo asteroids are both closely similar and comparatively flat. In addition, we observe that the SFD of the non-halo asteroids is quite different to those of the family and halo asteroids (Supplementary Figure 5) suggesting that the non-halo asteroids originate from ghost families. These further arguments allow us to estimate that the family and halo asteroids constitute ∽ 85% of the asteroids in the IMB and originate from the Flora, Vesta, Nysa, Polana and Eulalia families, while the remaining non-halo asteroids constitute ∽ 15% of the IMB and originate possibly from the same major families but, more likely, given the observed change of mean $I$ with $H$ and the observed SFD, from a few ghost families.

These arguments lead us to two important conclusions that do not depend on whether the non-halo asteroids originate from the major families or from a few ghost families. We only require that all of the IMB asteroids originate from the disruption of a few large bodies and we have argued that the observed correlations of the mean orbital elements of the IMB asteroids with H support that argument. Firstly, we conclude that we must seek explanations for the differences between the various meteorite groups, and the near-Earth asteroids, in the evolutionary histories of a comparatively few, large, parent bodies[2]. Secondly, our argument that the asteroids in the IMB originate from the disruption of a few large bodies supports the model that all asteroids formed big through the gravitational collapse of material in a protoplanetary disk[3]. Finally, we note that our observation that the slopes of the SFDs of the major families in the IMB tend to zero for $H \sim 18$ (Figure 3c) suggests that the peak in the SFD of the IMB at $H \sim 18$ may be real and not simply an artifact of observational selection. While recognizing that the asteroid families must have different ages and that we must analyze the evolution of their SFDs individually, the observed peak at $H \sim 18$, if real, suggests that small IMB asteroids, including future NEAs, are being lost from the IMB at approximately the same rate as they are being created through the disruption of larger asteroids.

# References

1. Hirayama, K. Groups of asteroids probably of common origin. *Astron. J* . **31**, 185-188 (1918).

2. Burbine, T.H., McCoy, T.J., Meibom, A., Gladman, B. & Keil, K. Meteoritic parent bodies: Their number and identification. In *Asteroids III* (W. F. Bottke Jr., A. Cellino, P. Paolicchi & R. P. Binzel (eds)), University of Arizona Press, Tucson, 653-667 (2002).

3. Johansen, A., Oishi, J. S., Mac Low, M-M., Klahr, H., Henning, T. & Youdin, A. Rapid planetesimal formation in turbulent circumstellar disks. *Nature* **448**, 1022-1025 (2007).

4. Knežević Z. & Milani A. Synthetic proper elements for outer main belt asteroids. *Celest. Mech. Dyn. Astr.* **78**, 17-46 (2000).

5. Nesvorný, D. Nesvorný HCM Asteroid Families V3.0. EAR-A-VARGBDET-5-NESVORNYFAM-V3.0. NASA Planetary Data System (2015).

6. Nesvorný, D., Broz, M. & Carrruba, V. Identification and dynamical properties of asteroid families. In *Asteroids IV* (P. Michel, F. DeMeo, and W. F. Bottke, Eds). U. Arizona Press, 297-321 (2015).

7. Zappala, V., Cellino, A., Farinella, P. & Knežević, Z. Asteroid families. I. Identification by hierarchial clustering and reliability assessment. *Astron. J.* **100**, 2030-2046 (1990).

8. Parker, A., Ivezić, Z. Juric, M., Lupton, R., Sekora, M. D., & Kowalski, A. The size distributions of asteroid families in the SDSS Moving Object Catalog 4. *Icarus* **198**, 138-155 (2008).

9. Milani, A., Cellino, A., Knežević, Z., Novaković, B., Spoto, F. & Paolicchi, P. Asteroid families classification: Exploiting very large datasets. *Icarus* **239**, 46-73 (2014).

10. Bottke, W. F., Vokrouhlický, D., Rubincam, D. P. & Broz, M. The effect of Yarkovsky thermal forces on the dynamical evolution of asteroids and meteoroids. In *Asteroids III* (W. F. Bottke Jr., A. Cellino, P. Paolicchi, and R. P. Binzel (eds)), University of Arizona Press, Tucson, 395-408 (2002).

11. Spoto, F., Milani, A. & Knežević, Z. Asteroid family ages. *Icarus* **257**, 275-289 (2015).

12. Dohnanyi, J. S. Collisional Model of Asteroids and Their Debris. *J. Geophys. R.* **74**, 2531-2554 (1969).

13. Durda, D. D. & Dermott, S. F. The collisional evolution of the asteroid belt and its contribution to the zodiacal cloud. *Icarus* **130**, 140-164 (1997).

14. Vokrouhlický, D. & Farinella, P. Efficient delivery of meteorites to the Earth from a wide range of asteroid parent bodies. *Nature* **407**, 606-608 (2000).

15. Bottke, W. F., Rubincam, D. P. & Burns, J. A. Dynamical evolution of main belt meteoroids: Numerical simulations incorporating planetary perturbations and Yarkovsky thermal forces. *Icarus* **145**, 301-331 (2000).

16. Jacobson, S.A., Marzari, F., Rossi, A., Scheeres, D.J. & Davis, D.R. Effect of rotational disruption on the size–frequency distribution of the Main Belt asteroid population *Mon. Not. R. Astron. Soc.* **439**, L95–L99 (2014).

17. Morbidelli, A. & Nesvorný, D. Numerous weak resonances drive asteroids toward terrestrial planet's orbits. *Icarus* **139**, 295-308 (1999).

18. Minton, D.A. & Malhotra, R. Dynamical erosion of the asteroid belt and implications for large impacts in the inner solar system. *Icarus* **207**, 744-757 (2010).

19. Delbo' M., Walsh K., Bolin B., Avdellidou C. & Morbidelli, M. Identification of a primordial asteroid family constrains the original planetesimal population. *Science* **357**, 1026-1029 (2017).

20. Murray, C. D. & Dermott, S. F. *Solar System Dynamics*. Cambridge Univ. Press (1999).

21. Masiero, J. R., Mainzer, A. K., Bauer, J. M., Grav, T., Nugent, C. R. & Stevenson, R. Asteroid Family Identification Using the Hierarchical Clustering Method and WISE/NEOWISE Physical Properties. *Astrophys. J* . **770**, 7-29 (2013).

22. Milani, A. & Nobili, A. Integration error over a very long time span. *Celest. Mech. Dyn. Astr.* **43**, 1–34 (1988).

23. The OrbFit Software Package. URL: http://adams.dm.unipi.it/orbfit/

24. Carruba, V., Burns, J. A., Bottke, W. & Nesvorný, D. Orbital evolution of the Gefion and Adeona asteroid families: close encounters with massive asteroids and the Yarkovsky effect. *Icarus* **162**, 308–327

(2003).

25. Fisher, R. A. *Statistical Methods for Research Workers*. 24th edition. Hafner, Darien, Connecticut (1970).

**Acknowledgements** A.A.C. wishes to thank Prof. Zoran Knežević for his valuable advice on using $orbit9$ and on generating proper elements and to acknowledge the SFI/HEA Irish Centre for High-End Computing (ICHEC) for the provision of computational facilities and support. Astronomical research at the Armagh Observatory and Planetarium is funded by the Northern Ireland Department of Communities (DfC).

**Author Information** Reprints and permissions information is available at www.nature.com/reprints.

**Competing financial interests** The authors declare no competing financial interests.

**Corresponding authors** Correspondence and requests for materials should be addressed to sdermott@ufl.edu.

**Author Contributions** S.F.D. initiated and directed the research and wrote the paper. A.A.C. performed the numerical investigations of chaotic orbital evolution and wrote the corresponding part of the methods section. D.L., T.J.J.K. and J.M.R. contributed to the data analysis.

**Fig. 1: IMB Asteroid Families. a,** Proper $I$ and **b,** proper $e$ are plotted against orbital frequency for asteroids in the inner main belt (IMB) with absolute magnitude $H < 16.5$. The major asteroid families are shown colour-coded: Nysa-Polana-Eulalia (magenta), Massalia (blue), Baptistina (yellow), Flora (cyan) and Vesta (green). Black points are the non-family asteroids defined by Nesvorný[5,6]. The green shaded region in **b** is the Mars-crossing zone and the dashed curve in **a** is the $\nu_6$ secular resonance[20]. In the green shaded region (and above) the orbit of an asteroid can cross that of Mars. The width of this zone depends on the orbital eccentricity of Mars which varies from 0 to ~0.14 on a 2 $My$ timescale[20] (the upper bound of the zone corresponds to Mars in a circular orbit). **c,** A scatter plot of the proper $e$ and proper $I$. **d,** The non-family asteroids defined by Nesvorný[5,6] with points colour-coded according to the maximum Lyapunov Characteristic Exponent, mLCE, calculated by Knežević and Milani[4] with red being the most chaotic (mLCE>0.00012 year[-1]) and black the most stable (mLCE<0.00012 year[-1]).

**Fig. 2: Dependence of the mean eccentricity and the mean inclination on asteroid size. a,** Changes of the mean proper $I$ and **b,** the mean proper $e$ with absolute magnitude, $H$, for all family (yellow) and non-family (black) asteroids in the inner main belt (IMB) and for the Nysa-Polana-Eulalia (magenta), Massalia (blue), Vesta (green) and Flora (cyan) families. Each point represents an average value in a bin of width $dH = 0.5$. The statistical significances of the observed correlations are given in Supplementary Table 2.

**Fig. 3: Asteroid size-frequency distributions. a,** Histograms of the change with absolute magnitude $H$ of the number of asteroids $dN$ in a box of width $dH = 0.1$. The plots show all asteroids in the IMB (green), the non-family asteroids (black) and the all family asteroids (magenta). Nominal asteroid diameters have been estimated from $H$ assuming an albedo of 0.15 for all asteroids; a value that reflects the contributions from both the low albedo ($< 0.1$) and the high albedo ($> 0.2$) asteroids within the most populous IMB families[21]. **b,** Cubic polynomial fits to the histograms shown in **a** for asteroids with $16.5 < H < 14$ and similar histograms for the asteroids in the separate families with Poisson error bars determined by the numbers in the boxes. **c,** Variation with absolute magnitude $H$ of the slopes to the cubic polynomial fits shown in **b.** The error bars on the slopes were determined using Monte Carlo simulations given the data points and their 1-sigma uncertainties shown in **b. d,** Ratios of the number of asteroids $dN$ in a box of width $dH = 0.1$ for the major (most numerous) families.

**Fig. 4: Chaotic orbital evolution of resonant and non-resonant asteroids. a,** Evolution of the standard deviation (diamonds) of the proper eccentricity $e_p$ and the proper inclination $I_p$ for two sets of resonant test particles started in the 1M:2A mean motion resonance and integrated for $10^8 \, yr$ with $e_p = 0.15$ and $e_p = 0.19$ respectively and with $I_p = 2.5$ deg in both cases; $M$ and $A$ denote the mean motions of Mars and an asteroid. The respective values of the standard deviations reached at the end of the simulations were $\sigma(e_p) = 0.014$ and $\sigma(e_p) = 0.011$, and $\sigma(I_p) = 0.24 \, deg$ and $\sigma(I_p) = 0.21 \, deg$. Also shown are the same quantities (squares) for a third set of non-resonant

clones started with $e_p = 0.2$ and $I_p = 2.5\ deg$ but outside of any resonance and integrated for $7.2\ 10^7 yr$. Here the final values were $\sigma(e_p) = 0.0032$ and $\sigma(I_p) = 0.067\ deg$. The red dashed lines for the resonant clones represent fits to the data segments from $t = 10^7 yr$ until $t = 10^8 yr$. For the non-resonant clones the same fit (blue dashed line) was applied to a shorter segment of data, starting from t= $3\ 10^7 yr$. The fitted functions were of the form $log_{10}\ \sigma(t) = c + b log_{10}(t)$. For the eccentricity increases of the resonant groups: for $e_p = 0.15$, $b = 0.423$ and $c = -5.25$; for $e_p = 0.19, b = 0.422$ and $c = -4.35$. For the non-resonant group, $b = 0.145$ and $c = -3.68$. For the inclination increases of the resonant groups, the respective values are: for $e_p = 0.15, b = 0.429$ and $c = -4.07$; for $e_p = 0.19, b = 0.472$ and $c = -4.49$. For the non-resonant group, $b = 0.290$ and $c = -3.50$.

**METHODS**

**Estimates of chaotic orbital evolution timescales.** To determine how the orbits diffuse over time, we generated populations of test particles with common proper elements (see below) both within and outside the 1M:2A mean motion resonance, where M and A denote the mean motions of Mars and an asteroid. All particles shared a common proper inclination, $I_p$= 2.5 deg. The resonant population is made up of two groups of 400 particles each, one group with $e_p = 0.15$ and the other with $e_p = 0.19$. The non-resonant population, also of 400 particles, was assigned $e_p = 0.20$. The proper longitudes of perihelion and the ascending node were assigned uniformly random values. The initial osculating elements were formed by adding the chosen proper elements vectorially to the forced component (see below). The mean anomaly was randomised and the semimajor axis held fixed to the value $a = 2.4183440682158213\ au$ (resonant groups) and $a = 2.4230444551325832\ au$ (non-resonant group). The particles were integrated using the orbit9 code[22] available within the orbfit package[23] for $10^8 yr$ from the present with a maximum step size of $0.2\ yr$ and an output step of $10^4 yr$. The solar system model included the seven planets from Venus to Neptune. The mass of Mercury was added to that of the sun; the initial planetary state vectors were taken from the JPL DE405 ephemeris at the epoch JD2457000.5 and referred to the Sun-Mercury barycenter.

The cumulative effect of chaotic orbital evolution of the orbits can be quantified through the function[24]

$$V_c(t) = \frac{1}{N-1} \sum_{i=0}^{N} [c_i(t) - c_i(0)]^2 \qquad (2)$$

where $i$ runs through the particles for each group, $c$ can be $a_p, e_p$ or $I_p$ and $N$ is the sample size. The sample standard deviation is then $\sigma_c(t) = \sqrt{V_c(t)}$. We calculate the time series $c_i(t)$ by averaging the integration output every 32 samples or $3.2 \times 10^5 yr, > 2 \times$ the periods of the principal harmonics in the output.

**Creating synthetic populations of asteroids with common proper elements.** For each test particle, the semimajor axis $a$, the eccentricity $e$, and the inclination $I$ as well as the two angles $\varpi$ and $\Omega$ that define the orientation of the orbit osculate due to the gravitational pull of the planets[20]. Apart from the semimajor axis, the variations can be thought of as the vector sums of the respective forced and proper vectors. For example, for the eccentricity

$$\begin{pmatrix} e \cos \varpi \\ e \sin \varpi \end{pmatrix} = \begin{pmatrix} e_f \cos \varpi_f \\ e_f \sin \varpi_f \end{pmatrix} + \begin{pmatrix} e_p \cos \varpi_p \\ e_p \sin \varpi_p \end{pmatrix} \qquad (3)$$

where the forced component (subscript $f$) depends primarily on the semimajor axes of the asteroid and the planets and on the planetary masses, while the proper component (subscript $p$) has a constant modulus that is intrinsic to that asteroid. The proper semimajor axis $a_p$ or, equivalently, the proper mean motion $n_p$ from Kepler's third law[20] is also constant and may be approximated by the long-term average of the semimajor axis $a$.

To generate test particle orbits with a given proper eccentricity and/or inclination for the simulations we first determine the respective forced vector at or near the starting epoch. This vector is common to all asteroids with the same $a$. We do this by averaging the osculating element vectors for a large enough set of asteroids. We describe here the procedure for the eccentricity. We identified 14,653 numbered asteroids[4] with proper mean motion $n_p$ between 94.5 and 96.5 $deg/yr$ and calculated the common forced vector for this set in two different ways:

(1) Processing the asteroids' osculating elements[20] using tools within Orbfit to calculate their proper elements. We then estimate the forced vector for each individual asteroid as

$$\begin{pmatrix} e_f \cos \varpi_f \\ e_f \sin \varpi_f \end{pmatrix} = \begin{pmatrix} e \cos \varpi \\ \mathrm{e} \sin \varpi \end{pmatrix} - \begin{pmatrix} e_p \cos \varpi_p \\ e_p \sin \varpi_p \end{pmatrix} \qquad (4)$$

The common force vector is the geometric average $< (e \cos \varpi, \mathrm{e} \sin \varpi \,) >$ of these vectors for all the asteroids.

(2) Forming the osculating vector for each asteroid and taking the average $< (e \cos \varpi, \mathrm{e} \sin \varpi \,) >$. If the azimuth of the proper vectors is randomly distributed, then averaging estimates the common forced vector. There is good agreement between the two methods, the results differing by $2 \times 10^{-3}$ or ~5% for the eccentricity vector and 0.1 $deg$ or ~10% for the inclination vector. Since the second method only requires a priori knowledge of the osculating elements, we adopted it for the work reported here.

**Statistics of correlations.** The correlations between mean proper $e$ and mean proper $I$ and $H$ shown in Figure 2 and in Supplementary Figures 2 and 4 used binned data. Here we estimate the statistical significance of these correlations by calculating the linear correlation coefficient $r$ for the individual data points. The statistical significance of $r$ can be estimated from

$$t = |r|\sqrt{(N-2)/(1-r^2)} \qquad (5)$$

where $t$ is Student's statistic[25]. If the number of data points N is large then the distribution of $t$ can be assumed to be normal and $t$ then represents the number of standard deviations from the mean. In Supplementary Table 2 we use the proper orbital elements and absolute magnitudes of the numbered asteroids in the Knežević and Milani[4] 2017 database with the restrictions $2.1 < a < 2.5\ au, e < 0.325, I < 16.0\ deg$ and $16.5 > H > 13.5$.

**Analysis of changes of mean $e$ and $I$ with $H$ in subsections of the IMB.** The asteroids in the main belt are usually divided into three groups (inner, middle and outer) as determined by the locations of the strongest Jovian mean motion resonances or Kirkwood gaps. In this paper we have treated all asteroids in the IMB as one group, but there are advantages to dividing these asteroids into further subgroups. The major theme of this paper is that a large fraction of the non-family asteroids in the IMB originate from the major families, but the semimajor axis ranges of these families are less than the width of the IMB and a more natural length scale to use would be one determined by the semimajor axis ranges of the families (Supplementary Table 1).

The IMB has five major families: Flora, Vesta and the Nysa-Polana-Eulalia complex. Here we divide the 2017 numbered non-family asteroids with $e < 0.325$ and $I < 16\ deg$ into three ranges.

(1) $2.14 < a < 2.24\ au$. The only major family in this range is the Flora family and there are no asteroids from the Vesta family or the Nysa-Polana-Eulalia complex. The number of non-family asteroids in this set with $16.5 > H > 13.5$ is 5,638.

(2) $2.24 < a < 2.41\ au$. In this range there are asteroids from all five major families: Flora, Vesta and the Nysa-Polana-Eulalia complex. The number of non-family asteroids in this set with $16.5 > H > 13.5$ is 20,629.

(3) $2.41 < a < 2.50\ au$. In this range there are only asteroids from the Vesta family and the Nysa-Polana-Eulalia complex and none from the Flora family. The number of non-family asteroids in this set with $16.5 > H > 13.5$ is 7,908.

Supplementary Figure 2a shows that the increase in mean $I$ with decreasing $H$ that we observe for the non-family asteroids in the IMB as a whole (Figure 2) derives from those asteroids in the outer two ranges of the IMB. These two ranges are dominated by asteroids in the Vesta family and the Nysa-Polana-Eulalia complex. For asteroids in the inner range of the IMB, in which the Flora family is the only major family, in contrast to the outer ranges we observe a decrease in mean $I$ with decreasing $H$. The Flora family has an SFD that is quite different from those of the Vesta family and the Nysa-Polana-Eulalia complex (Figure 3b) and the differences in the variations of mean $I$ with $H$ are, perhaps, to be expected. However, given that the Flora family is the only known major family in the inner range of the IMB, the fact that we observe a decrease in mean $I$ with decreasing $H$ suggests that not all of the non-family asteroids can originate from the Flora family alone and, consequently, that some of the non-family asteroids in that semimajor axis range originate from one of more ghost families.

The distribution of proper $e$ and proper $I$ shown in Figure 1d strongly suggests that the major families are surrounded by halos of non-family asteroids that originate, at least in part, from the major families. Here we separate these non-family asteroids into halo and non-halo asteroids. Our aim is not to refine the separation of family and halo asteroids, but to define a set of non-halo asteroids that are well separated from both the family and the halo asteroids. Supplementary Figure 3 shows the distributions of the proper $e$ and the proper $I$ of the family and non-family asteroids in the IMB. The distributions of these orbital elements for the family asteroids are clearly bimodal showing the dominance of a few major families. Supplementary Figure 3a shows that the distribution of the proper $e$ of the non-family asteroids is also bimodal supporting the suggestion that these two populations of family and non-family asteroids have, at least in part, a common origin. In Supplementary Figure 3a we designate asteroids with $0.07 < e < 0.23$ as family plus halo asteroids. In Supplementary Figure 3b we designate asteroids with $0 < I < 8 \ deg$ as family plus halo asteroids. All asteroids that do not belong to either of these two sets are designated as non-halo asteroids. This separation is a matter of judgment, but we have justified this judgment by showing that the halo and non-halo asteroids have quite different properties. With the restrictions $2.1 < a < 2.5 \ au, e < 0.325, I <$ 16.0 $deg$ and $16.5 > H > 13.5$: the total number of non-halo asteroids is 9,366 and the total number of family plus non-family asteroids is $27{,}267 + 34{,}331 = 61{,}598$.

**Data Availability Statement**

The data that support the plots within this paper and other findings of this study are available from the corresponding author upon reasonable request.

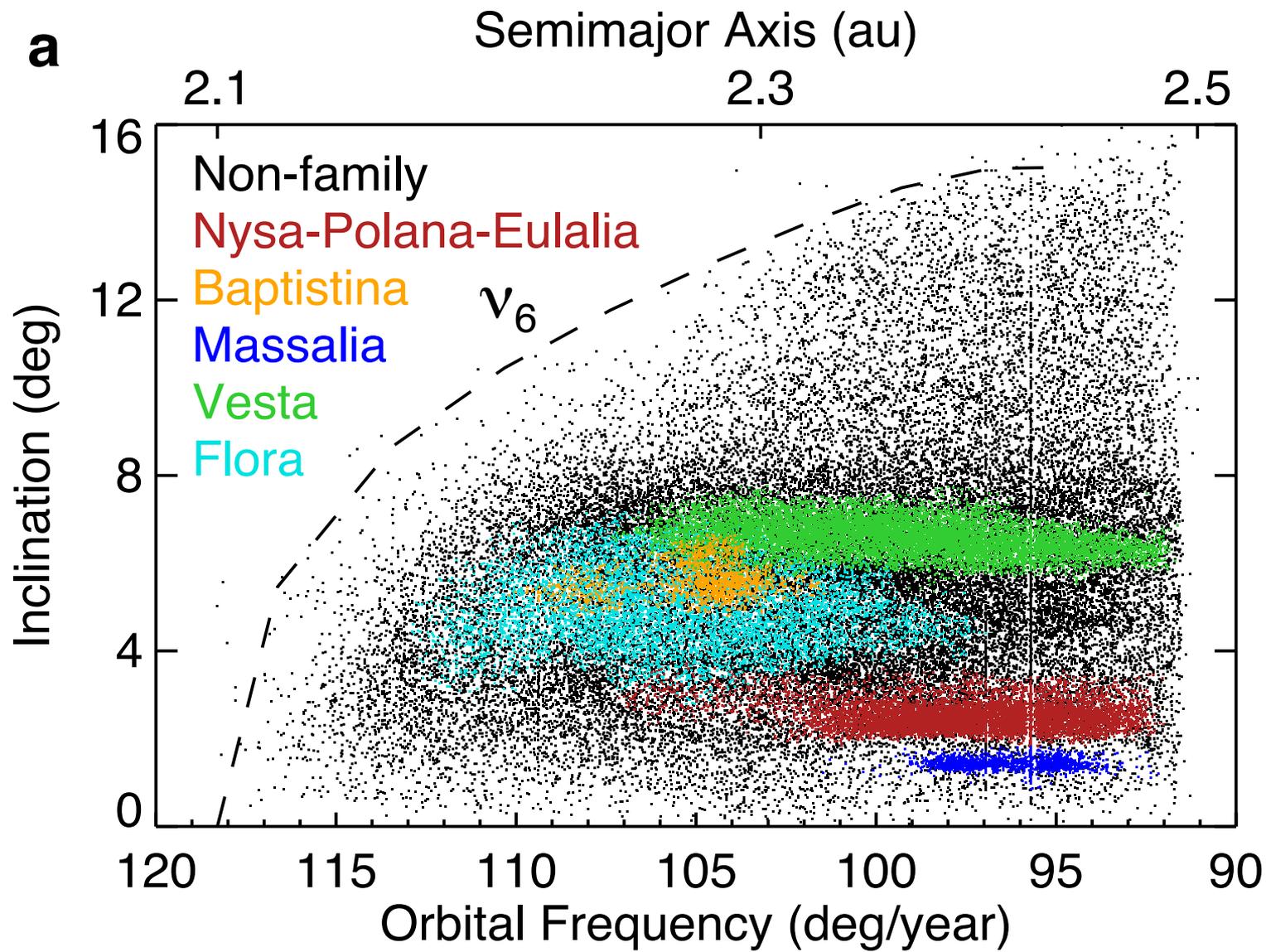

Figure 1a

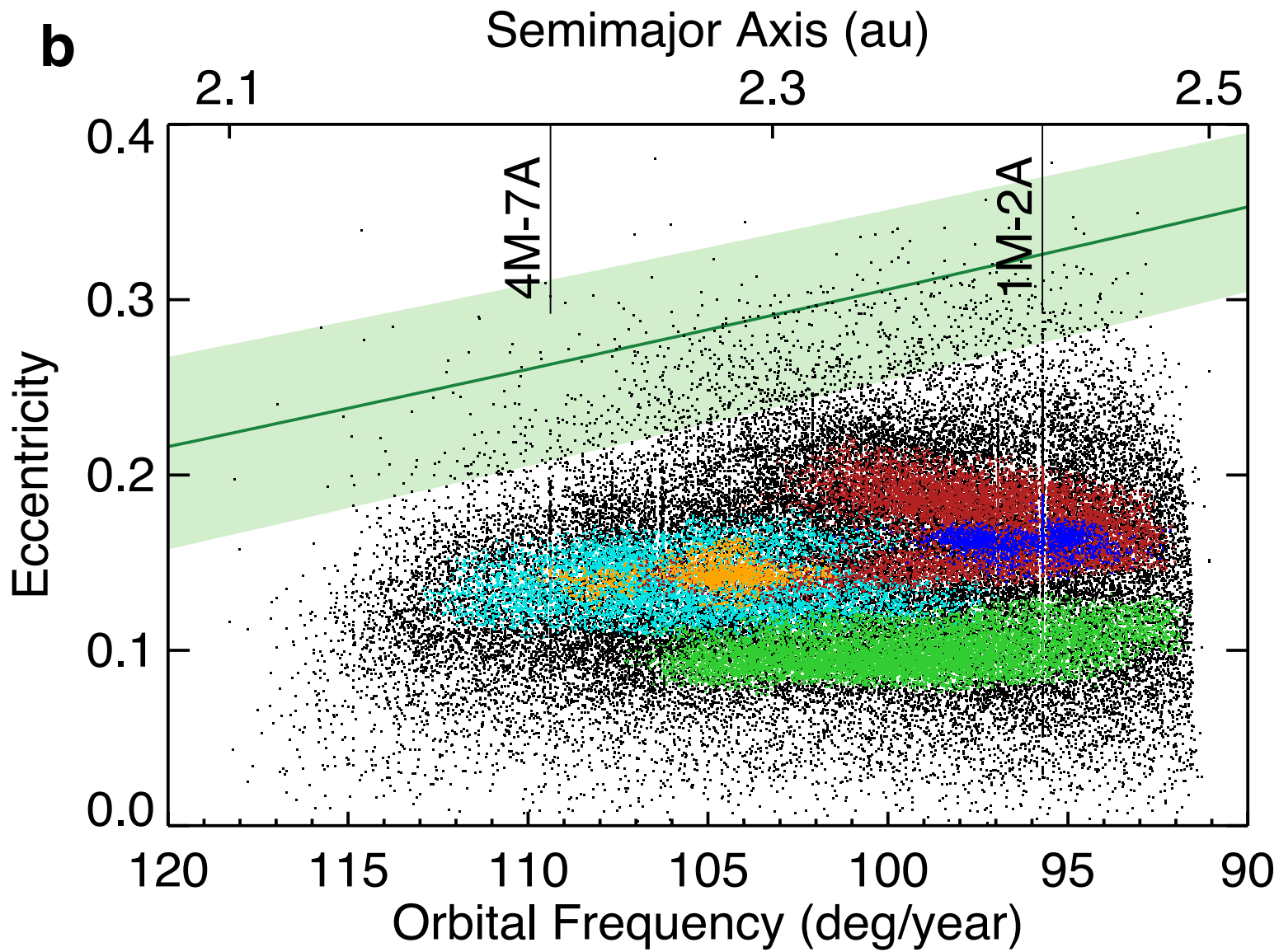

Figure 1b

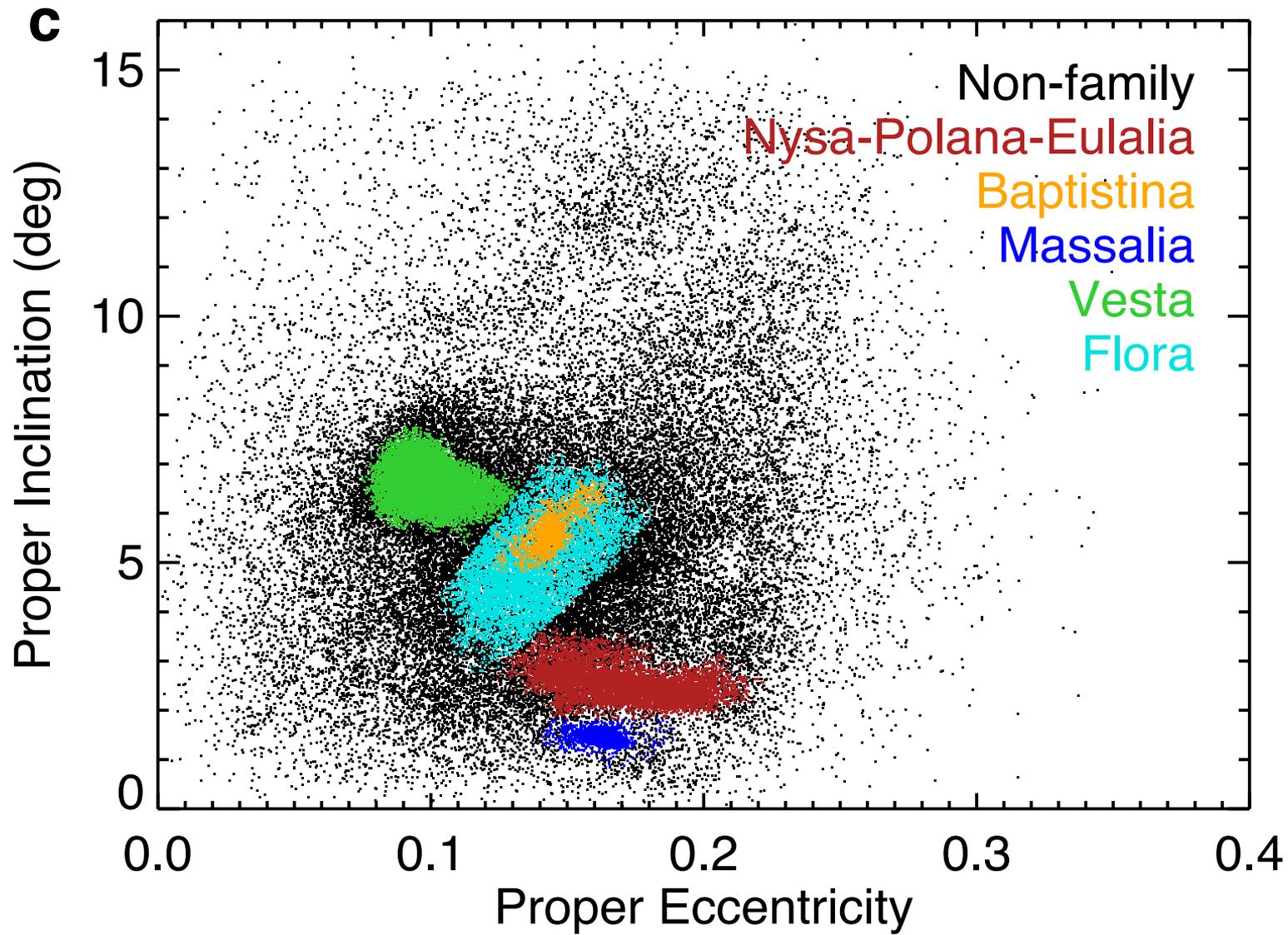

Figure 1c

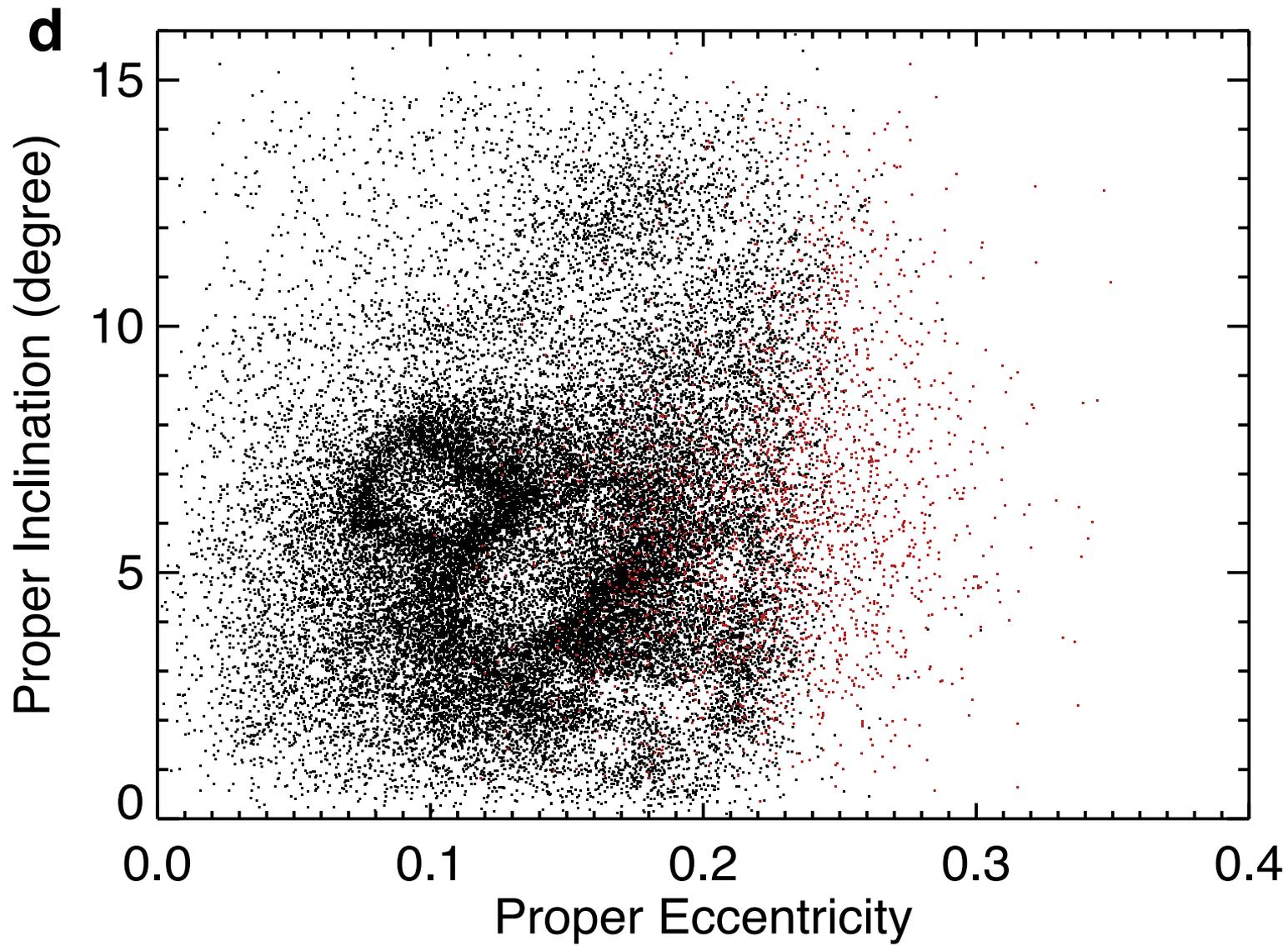

Figure 1d

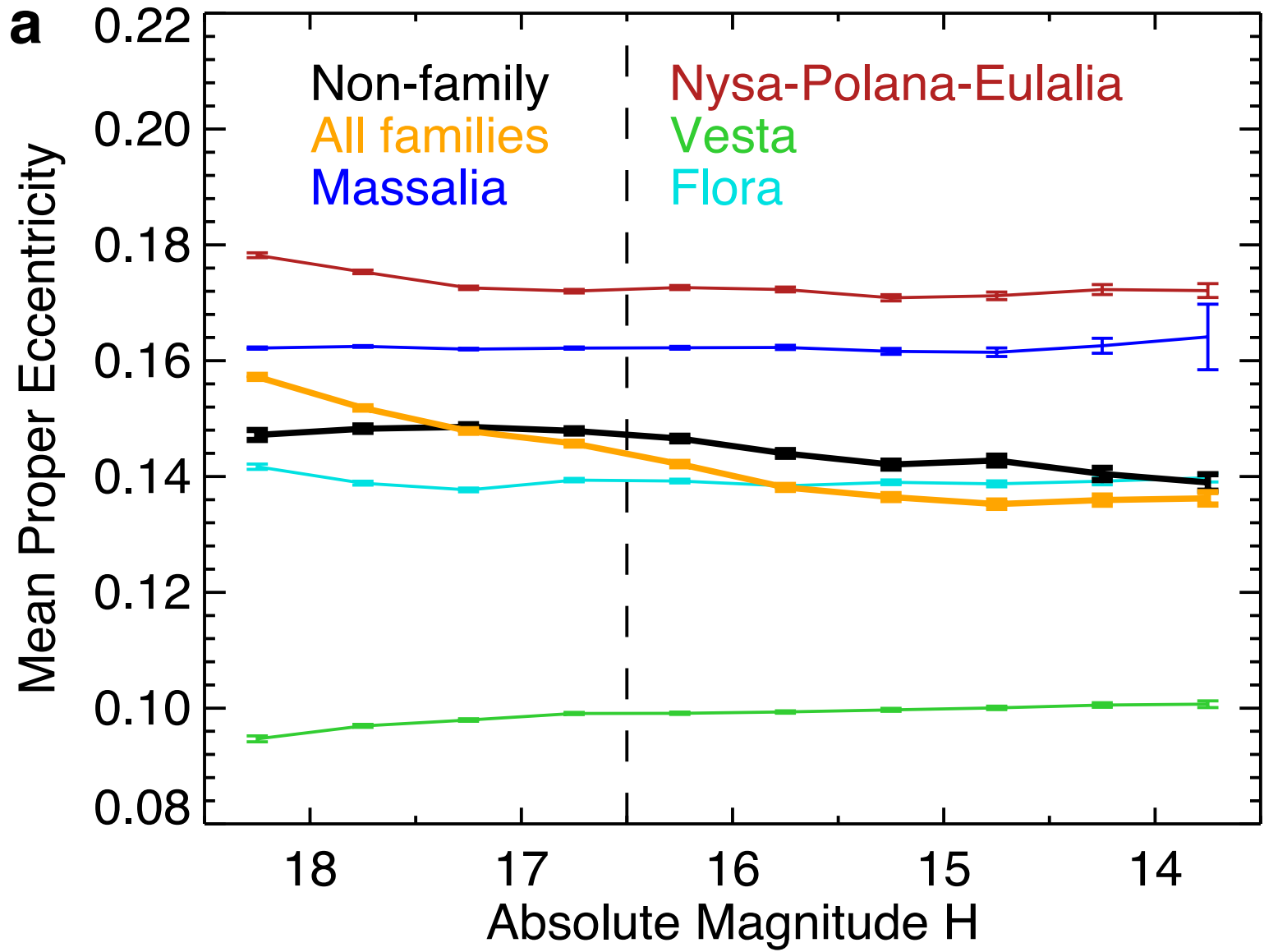

Figure 2a

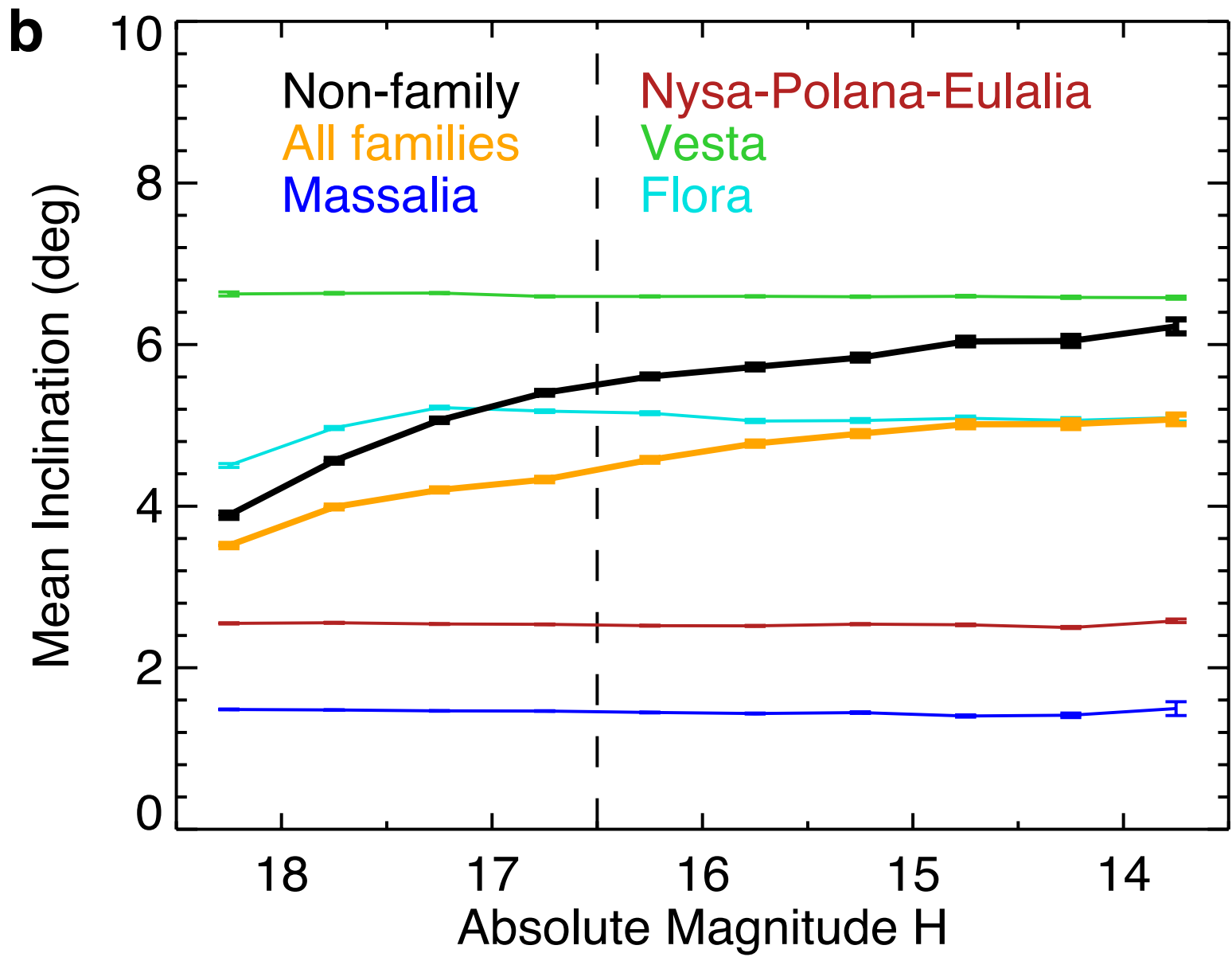

**b**

Non-family     Nysa-Polana-Eulalia
All families     Vesta
Massalia     Flora

Mean Inclination (deg)

Absolute Magnitude H

Figure 2b

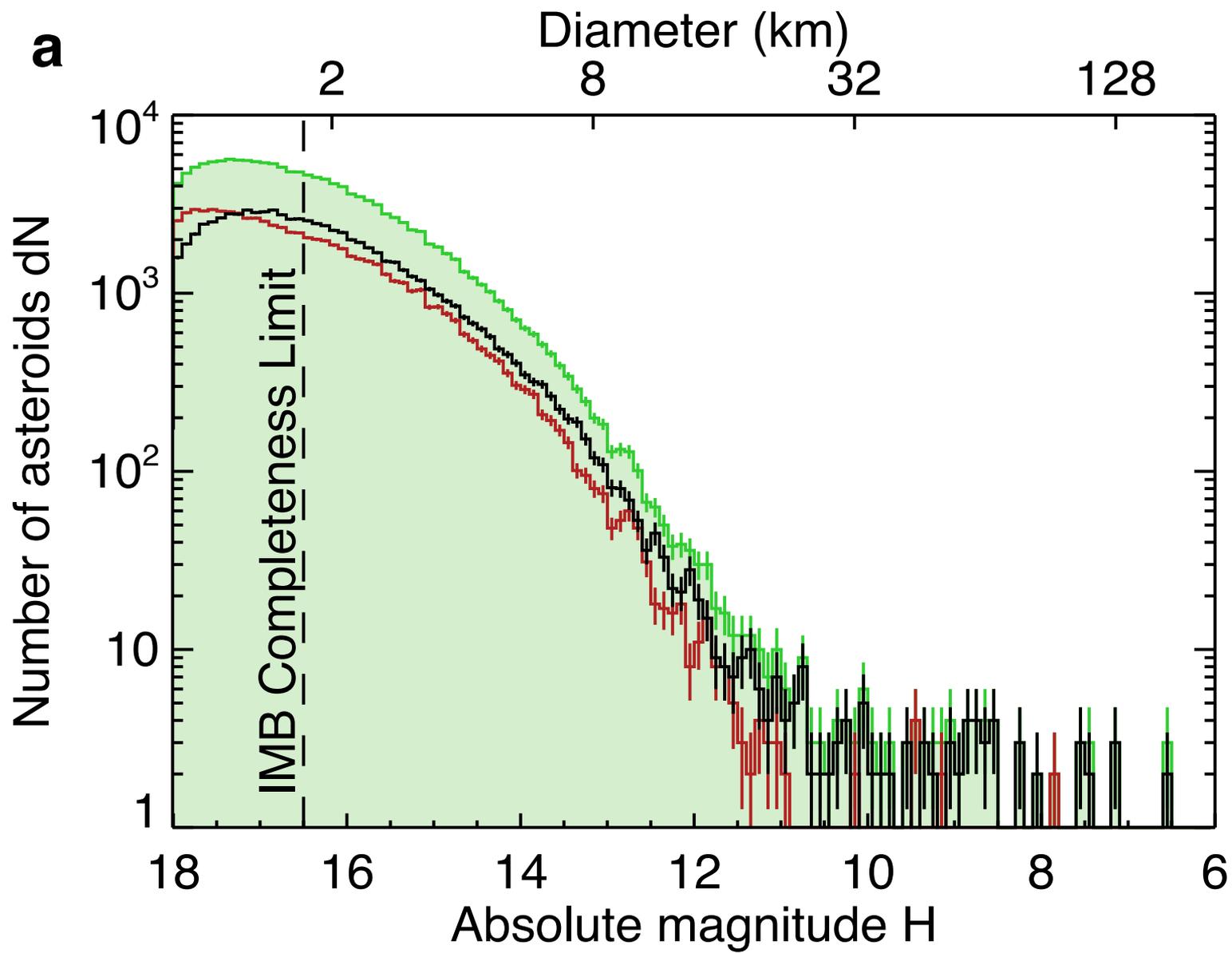

**a**

Figure 3a

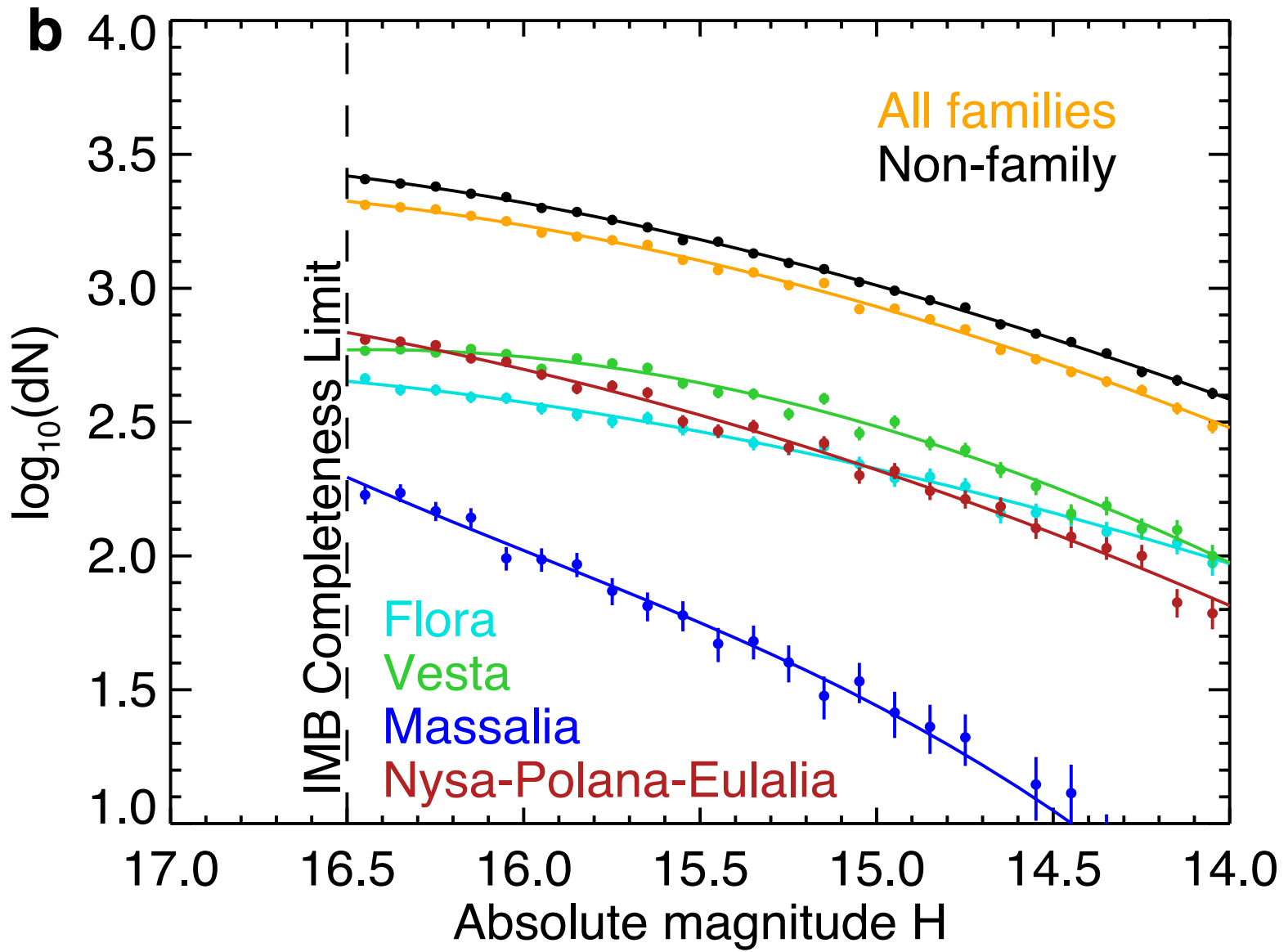

**b**

All families
Non-family

Flora
Vesta
Massalia
Nysa-Polana-Eulalia

IMB Completeness Limit

log₁₀(dN) — $\log_{10}(dN)$ (y-axis)

Absolute magnitude H (x-axis)

Figure 3b

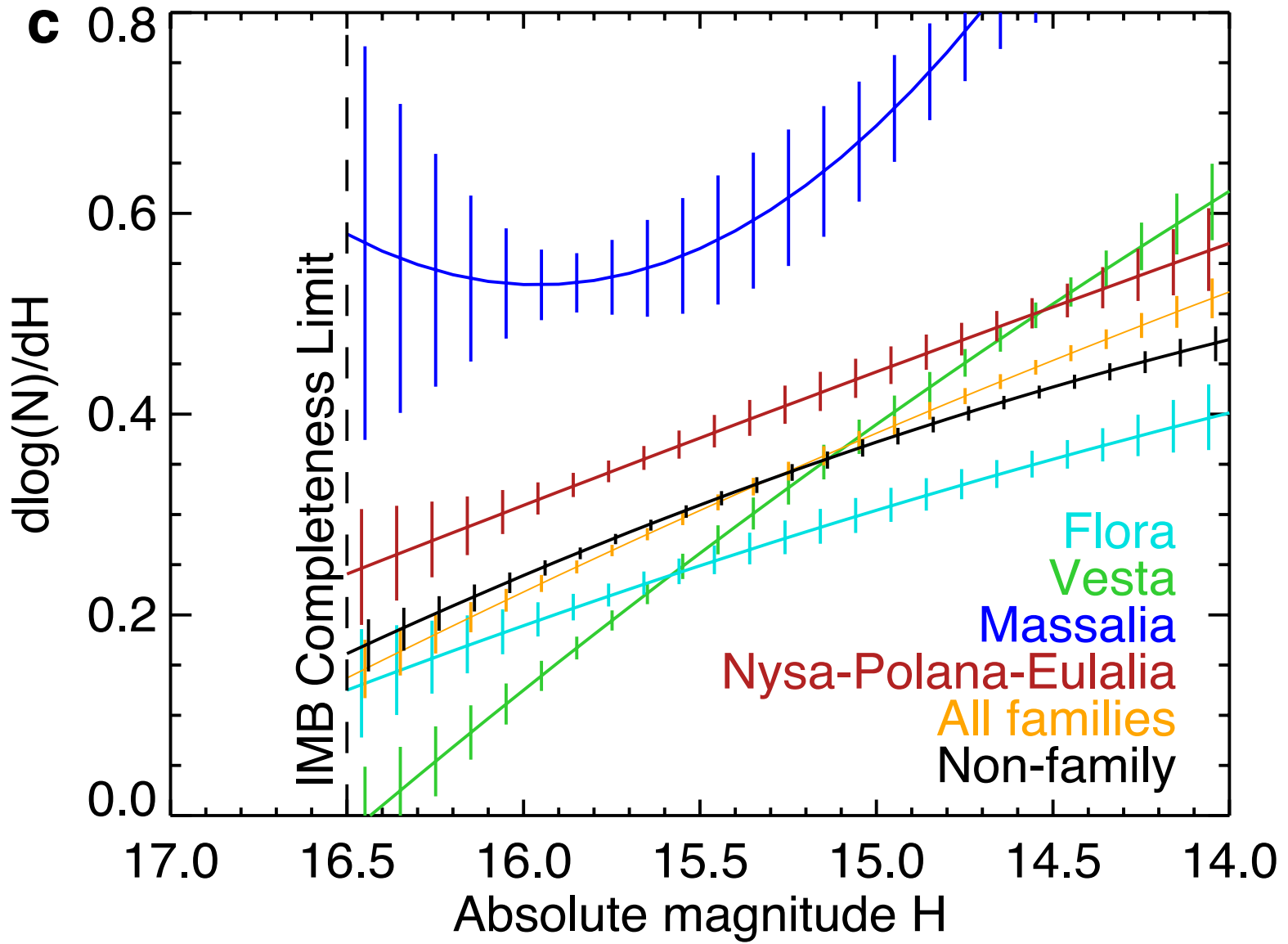

Figure 3c

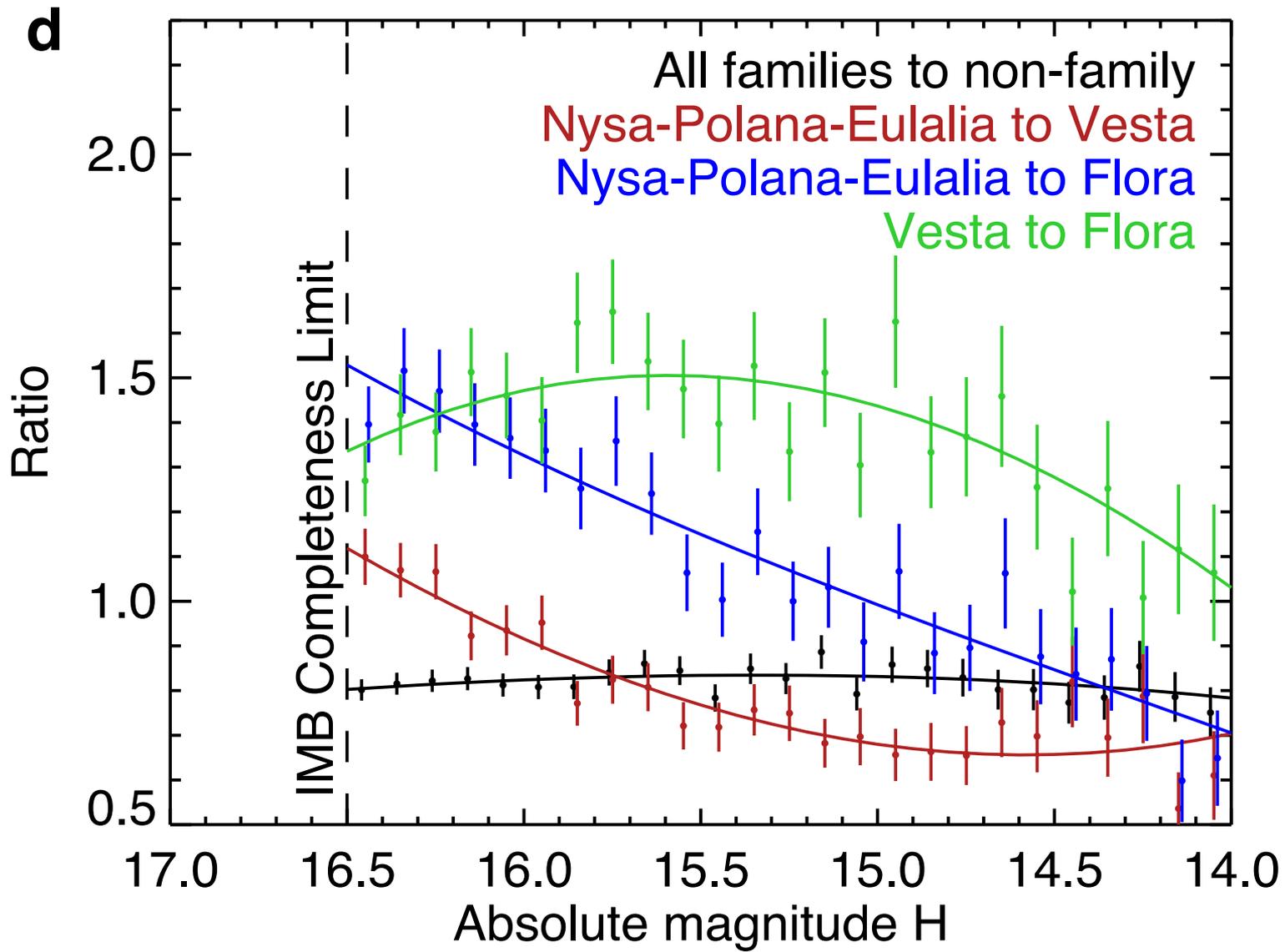

Figure 3d

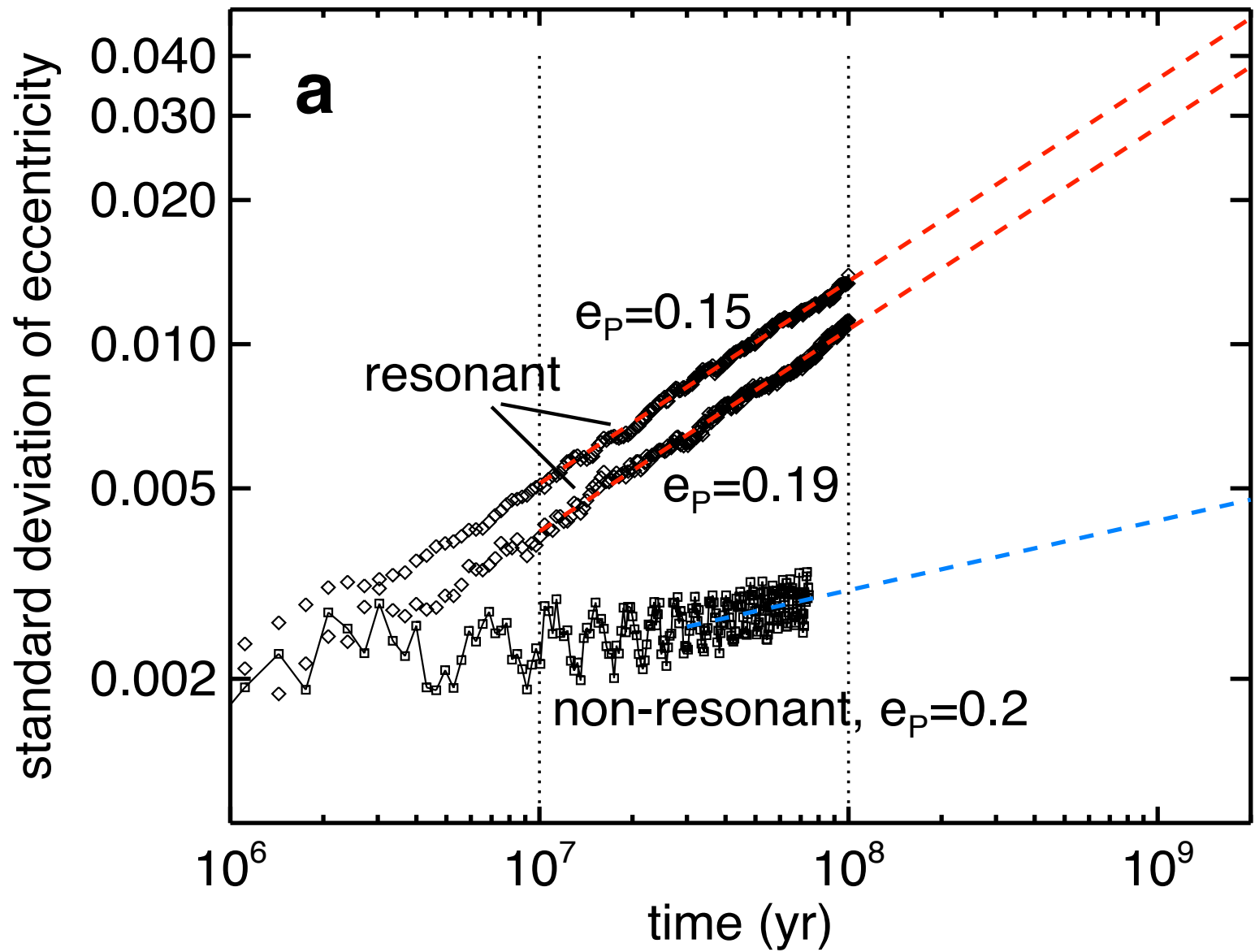

Figure 4a

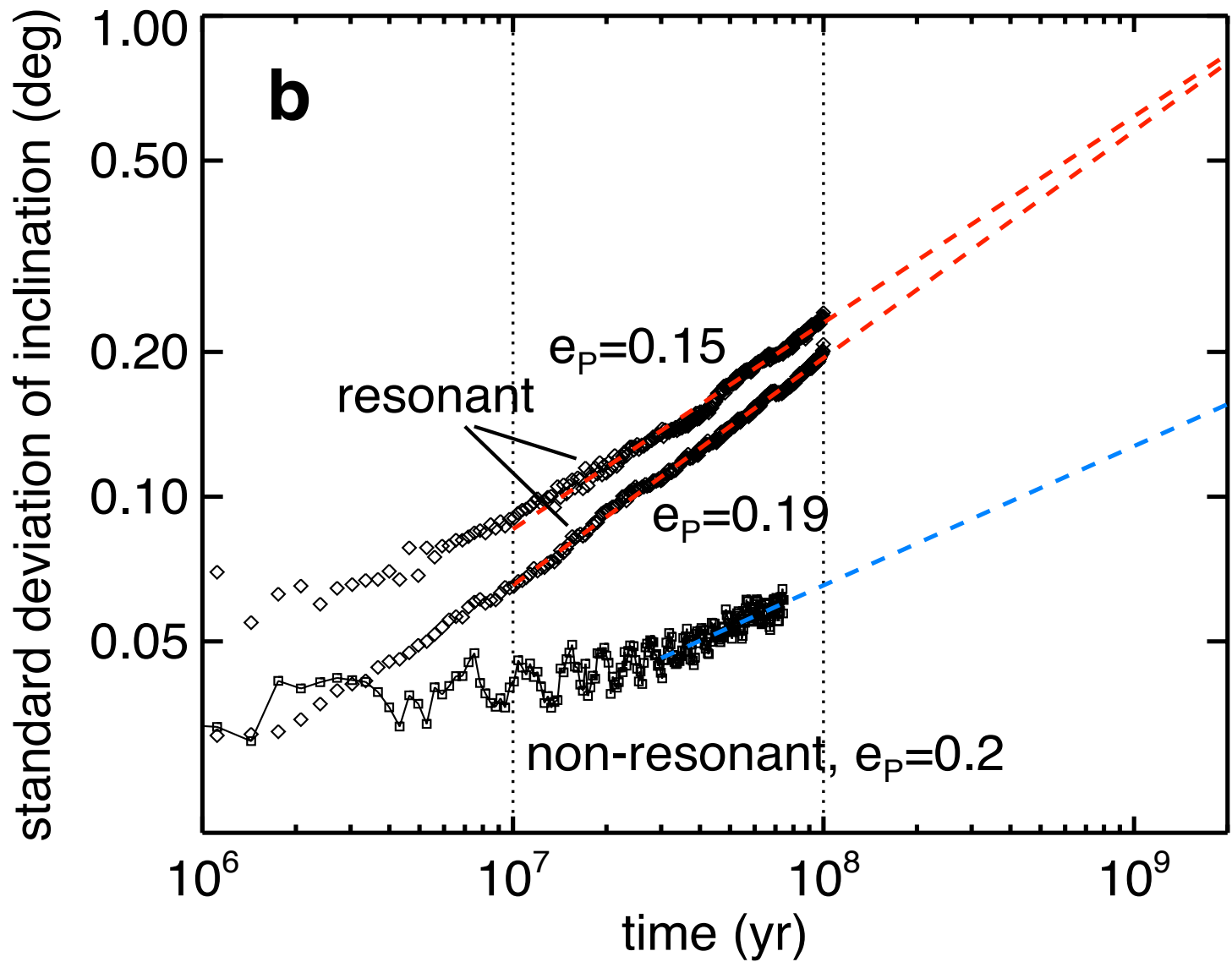

Figure 4b

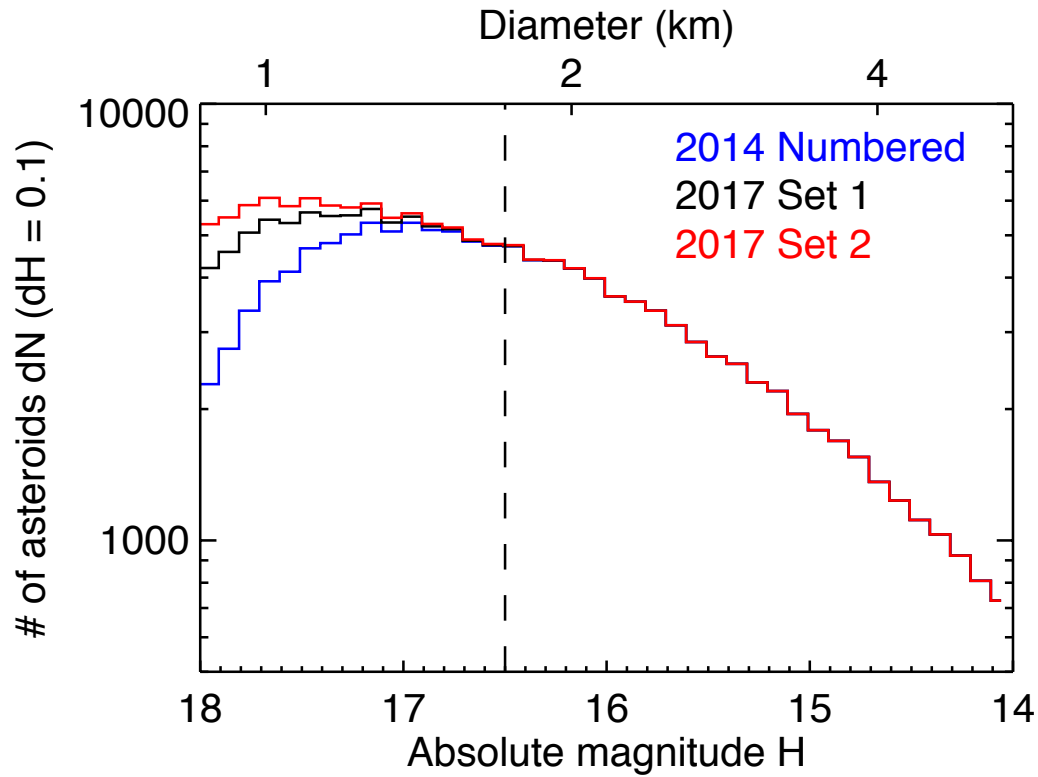

**Supplementary Figure 1: IMB completeness limit**

The catalog of synthetic proper elements for numbered asteroids calculated by Knežević and Milani[4] in April 2014 was superseded in June 2017 by a combined catalog containing both numbered and multi-opposition asteroids. (URL:http://hamilton.dm.unipi.it/~astdys2/propsynth/all.syn).

The three data sets shown above are from the 2014 catalog of numbered asteroids (green), the 2017 catalog of numbered asteroids (black), and the 2017 catalog of numbered plus multi-opposition asteroids (red).

With no restrictions on $e$ and $I$, but with $2.1 < a < 2.5\ au$ and $H < 16.5$, the number of asteroids, $N$, in the sets are:

2014 Numbered asteroids, $N = 66,435$
Set 1, 2017 Numbered asteroids, $N = 66,451$
Set 2, 2017 Numbered plus multi-opposition asteroids, $N = 66,494$

For $H < 16.5$, the differences between these data sets are negligible and we conclude that $H < 16.5$ is a conservative completeness limit for the IMB. The limit could be justifiably increased to $H < 17$, but that is not necessary for this paper. In all of our figures, we use the 2017 catalog of numbered asteroids and the family designations defined by Nesvorný[5].

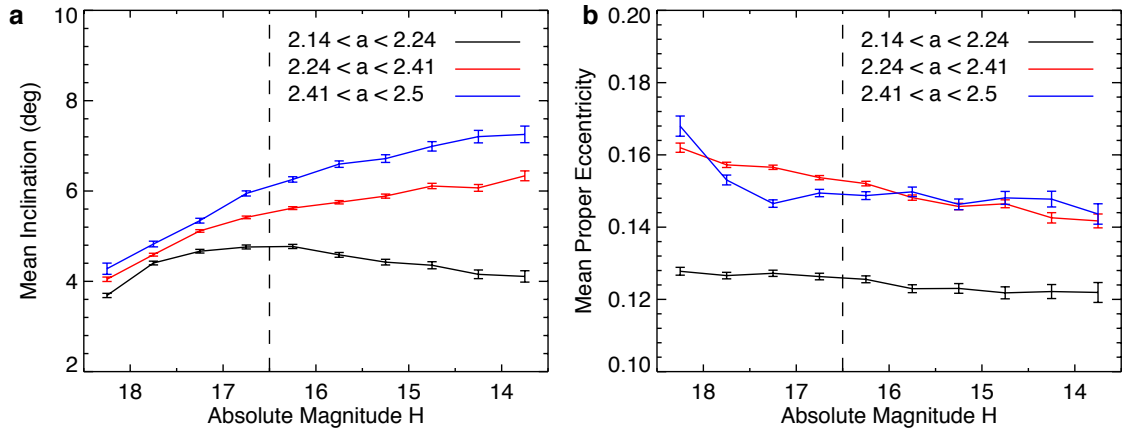

**Supplementary Figure 2: Three ranges of semimajor axis in the IMB**

The IMB has five major families: Flora, Vesta and the Nysa-Polana-Eulalia complex (Supplementary Table 1). Here we divide the 2017 numbered non-family asteroids with $e < 0.325$ and $I < 16\ deg$ into three ranges of semimajor axis.

$2.14 < a < 2.24\ au$. The only major family in this range is Flora and there are no asteroids from the Vesta or the Nysa-Polana-Eulalia families. The number of non-family asteroids in this set with $16.5 > H > 13.5$ is 5,638.

$2.24 < a < 2.41\ au$. In this range there are asteroids from all five major families: Flora, Vesta and the Nysa-Polana-Eulalia complex. The number of non-family asteroids in this set with $16.5 > H > 13.5$ is 20,629.

$2.41 < a < 2.50\ au$. In this range there are only asteroids from, the Vesta family and the Nysa-Polana-Eulalia complex and none from the Flora family. The number of non-family asteroids in this set with $16.5 > H > 13.5$ is 7,908.

The figures show data in bins of width $dH = 0.5$. Statistics on the observed linear correlations are shown in Supplementary Table 2.

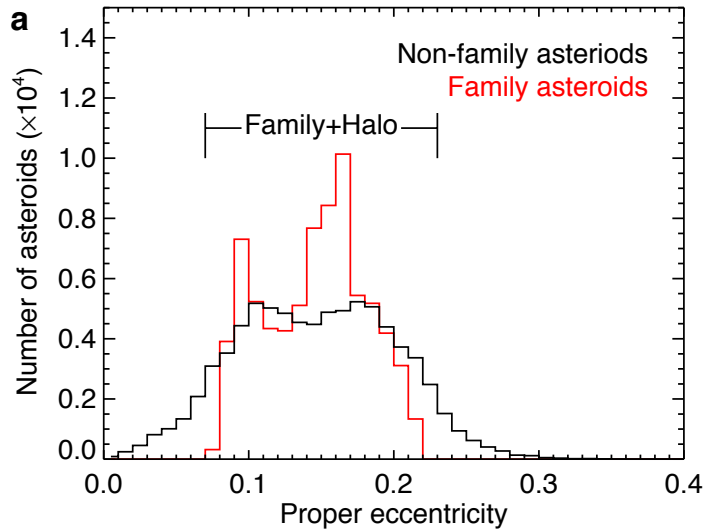

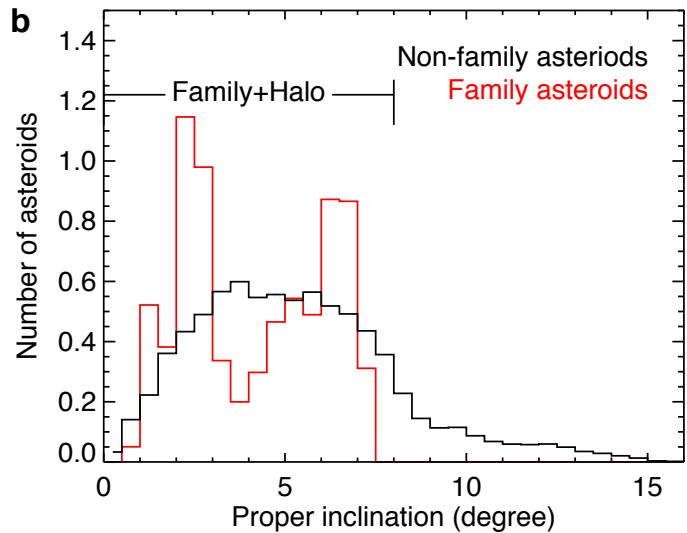

**Supplementary Figure 3: Definition of non-halo asteroids**

The purpose of this figure is not to separate the family and the halo asteroids, but to define a set of asteroids, the non-halo asteroids, that are well separated from both the family and the halo asteroids. In Supplementary Figure 3a we designate asteroids with $0.07 < e < 0.23$ as family and halo asteroids. In Supplementary Figure 3b we designate asteroids with $0 < I < 8$ $deg$ as family and halo asteroids. All asteroids that do not belong to either of these sets are designated as non-halo asteroids. With the restrictions: $2.1 < a < 2.5$ $au$, $e < 0.325$, $I < 16.0$ $deg$ and $16.5 > H > 13.5$; the total number of non-halo asteroids is 9,366 and the total number of family + non-family asteroids is $27,267 + 34,331 = 61,598$. This allows us to estimate that $\backsim$ **85**% of all the asteroids in the inner main belt originate from the Flora, Vesta, Nysa, Polana and Eulalia families with the remaining $\backsim$ **15**%, non-halo asteroids, originating from either the same families or, more likely, a few ghost families.

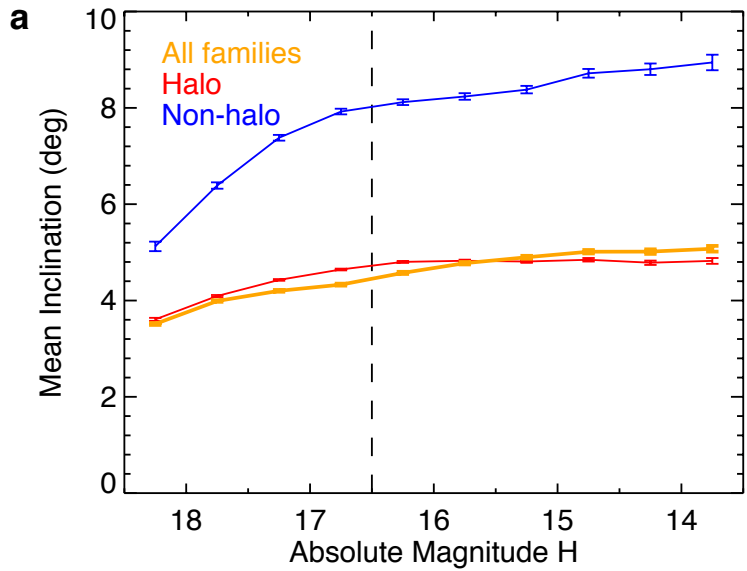

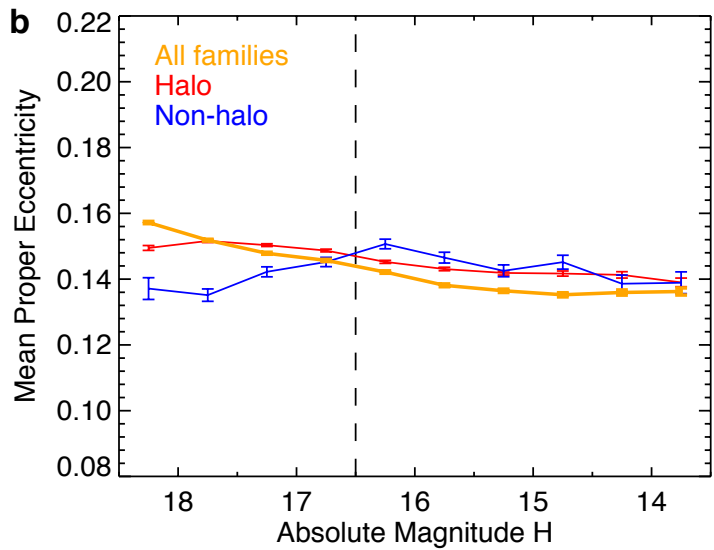

**Supplementary Figure 4: Correlations of mean *e* and mean *I* with *H* for halo and non-halo asteroids**

**a.** Variation of the mean proper inclination *I* with *H*.

**b.** Variation of the mean proper eccentricity *e* with *H*

The figures show data in bins of width *dH* = 0.5 for the family (yellow), the halo (red) and the non-halo asteroids (blue). Statistics on the observed correlations are given in Supplementary Table 2.

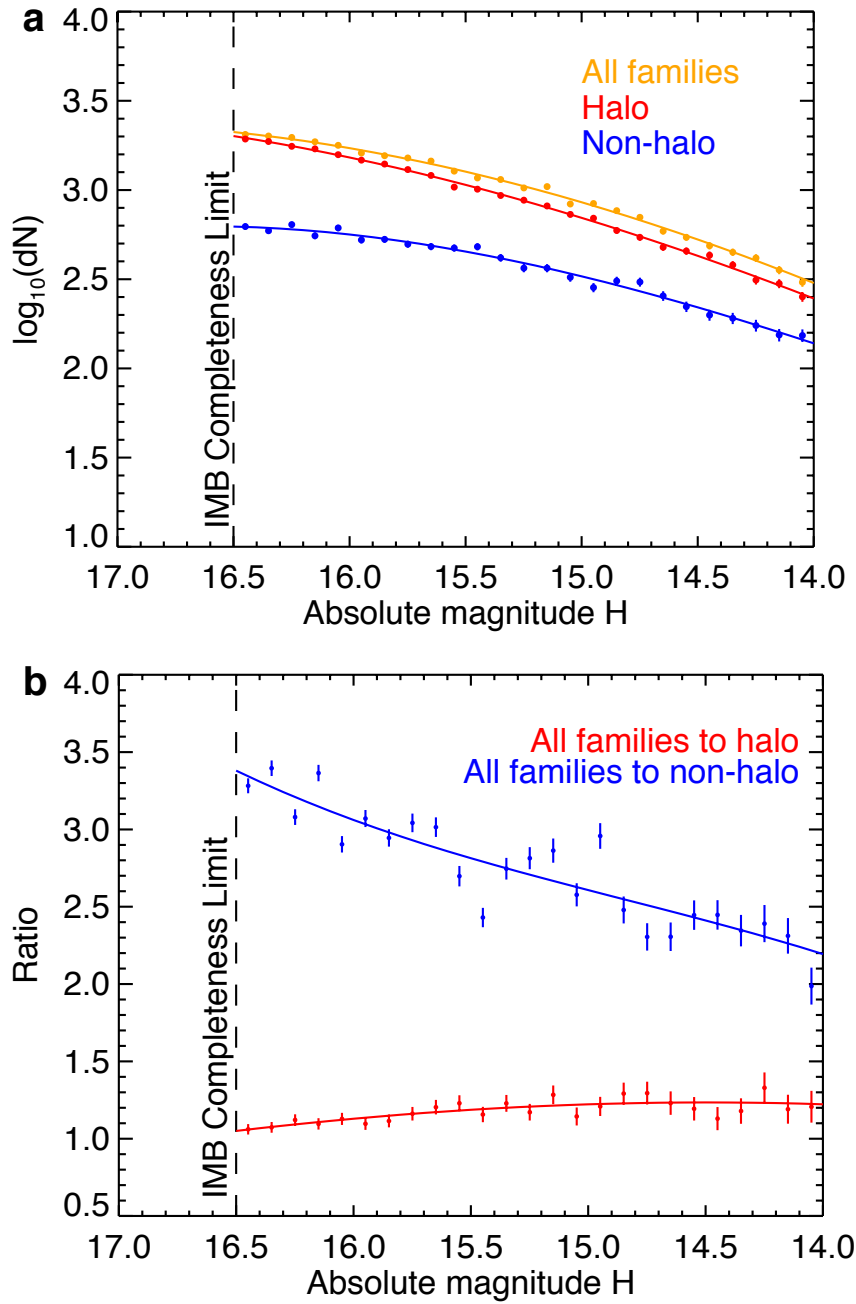

**Supplementary Figure 5: SFDs of halo and non-halo asteroids**

**a.** Cubic polynomial fits to the SFDs of the All family (yellow), halo (red) and non-halo (blue) asteroids with $16.5 < H < 14$.

**b.** Ratios of the number of asteroids $dN$ in a box of width $dH = 0.1$. The best fit curves for the variation of the ratio with $H$ is close to flat for the ratio of All families to halo asteroids (red), in contrast to the larger variations with $H$ observed for the ratio of All families to non-halo asteroids (blue).

**Supplementary Table 1. Nesvorný[5] families in the inner main belt**

| # | Family | | N | amin au | amax au | emin | emax | Imin deg | Imax deg |
|---|---|---|---|---|---|---|---|---|---|
| 401 | 4 | Vesta | 15,252 | 2.2407 | 2.4897 | 0.0746 | 0.1333 | 5.37 | 7.75 |
| 402 | 8 | Flora | 13,786 | 2.1624 | 2.4049 | 0.1034 | 0.1810 | 2.79 | 7.17 |
| 403 | 298 | Baptistina | 2,500 | 2.2010 | 2.3309 | 0.1223 | 0.1672 | 4.75 | 6.71 |
| 404 | 20 | Massalia | 6,424 | 2.3226 | 2.4801 | 0.1382 | 0.1935 | 0.84 | 1.82 |
| 405 | 44 | Nysa-Polana-Eulalia | 19,073 | 2.2412 | 2.4825 | 0.1245 | 0.2220 | 1.82 | 3.88 |
| | | | | | | | | | |
| 406 | 163 | Erigone | 1,776 | 2.3143 | 2.4224 | 0.1913 | 0.2218 | 4.34 | 5.85 |
| 407 | 302 | Clarissa | 179 | 2.3853 | 2.4206 | 0.1019 | 0.1124 | 3.20 | 3.56 |
| 408 | 752 | Sublimities | 303 | 2.4007 | 2.4882 | 0.0817 | 0.1031 | 4.79 | 5.81 |
| 409 | 1892 | Lucienne | 142 | 2.4270 | 2.4834 | 0.0842 | 0.1049 | 14.26 | 14.74 |
| 410 | 27 | Euterpe | 474 | 2.2969 | 2.4578 | 0.1731 | 0.2016 | 0.45 | 1.27 |
| 411 | 1270 | Datura | 6 | 2.2346 | 2.2349 | 0.1534 | 0.1535 | 5.30 | 5.30 |
| 412 | 21509 | Lucascavin | 3 | 2.2811 | 2.2812 | 0.1269 | 0.1269 | 5.23 | 5.23 |
| 413 | 84 | Klio | 330 | 2.2726 | 2.4627 | 0.1741 | 0.2114 | 8.29 | 11.18 |
| 414 | 623 | Chimaera | 108 | 2.3962 | 2.4887 | 0.1331 | 0.1625 | 14.12 | 15.36 |
| 415 | 313 | Chaldea | 132 | 2.3399 | 2.4718 | 0.2117 | 0.2437 | 9.79 | 12.03 |
| 416 | 329 | Svea | 48 | 2.4224 | 2.4895 | 0.0797 | 0.1046 | 15.76 | 16.38 |
| 417 | 108138 | 2001 GB11 | 9 | 2.4626 | 2.4669 | 0.1523 | 0.1527 | 3.92 | 3.94 |

Total N for families 401 through 417  = 60,545
Total N for families 401+402+405      = 48,111 = 79% of 60,545

79% of all Nesvorný family members in the IMB belong to the five most numerous families: Vesta, Flora, Nysa-Polana-Eulalia. There are ~2,000 asteroids with $H < 16.5$ interior to the IMB ($a < 2.1\ au$) in the Hungaria family.

## Supplementary Table 2. Statistics

In the tables shown below we use the proper orbital elements and absolute magnitudes of the numbered asteroids in the Knežević and Milani[4] database, with the restrictions that $2.1 < a < 2.5$ $au$, $e < 0.325$, $I < 16.0$ $deg$ and $16.5 > H > 13.5$.

### All families  (2.1<a<2.5 au)

|          |        | N      | \|r\|  | t    |
|----------|--------|--------|--------|------|
| Figure 2a | H v. I | 27,267 | 0.0775 | 12.8 |
| Figure 2b | H v. e | 27,267 | 0.0657 | 10.9 |

### Non-family  (2.1<a<2.5 au)

|          |        | N      | \|r\|  | t    |
|----------|--------|--------|--------|------|
| Figure 2a | H v. I | 34,331 | 0.0652 | 12.1 |
| Figure 2b | H v. e | 34,331 | 0.0429 | 8.0  |

### Non-family   (2.14<a<2.24 au)

|                      |        | N     | \|r\|  | t   |
|----------------------|--------|-------|--------|-----|
| Supplementary Figure 2a | H v. I | 5,638 | 0.1050 | 7.9 |
| Supplementary Figure 2b | H v. e | 5,638 | 0.0323 | 2.4 |

### Non-family   (2.24<a<2.41 au)

|                      |        | N      | \|r\|  | t    |
|----------------------|--------|--------|--------|------|
| Supplementary Figure 2a | H v. I | 20,629 | 0.0754 | 10.9 |
| Supplementary Figure 2b | H v. e | 20,629 | 0.0598 | 8.6  |

### Non-family   (2.41<a<2.50 au)

|                      |        | N     | \|r\|  | t   |
|----------------------|--------|-------|--------|-----|
| Supplementary Figure 2a | H v. I | 7,908 | 0.1028 | 9.2 |
| Supplementary Figure 2b | H v. e | 7,908 | 0.0208 | 1.8 |

### All families, Halo and Non-halo (2.1<a<2.5 au)

|                      |        |            | N      | \|r\|  | t    |
|----------------------|--------|------------|--------|--------|------|
| Supplementary Figure 4a | H v. I | All family | 27,267 | 0.0775 | 12.8 |
| Supplementary Figure 4a | H v. I | Halo       | 24,965 | 0.0027 | 0.4  |
| Supplementary Figure 4a | H v. I | Non-halo   | 9,366  | 0.0799 | 7.8  |

|                      |        |            | N      | \|r\|  | t    |
|----------------------|--------|------------|--------|--------|------|
| Supplementary Figure 4b | H v. e | All family | 27,267 | 0.0657 | 10.9 |
| Supplementary Figure 4b | H v. e | Halo       | 24,965 | 0.0424 | 6.7  |
| Supplementary Figure 4b | H v. e | Non-halo   | 9,366  | 0.0516 | 5.0  |